\documentclass[twocolumn,showpacs,preprintnumbers,amsmath,amssymb]{revtex4}%
\usepackage{graphicx}
\usepackage{dcolumn}
\usepackage{bm}
\usepackage{amsmath}
\usepackage{amsfonts}
\usepackage{amssymb}%
\providecommand{\U}[1]{\protect\rule{.1in}{.1in}}
\begin{document}
%

%TCIMACRO{\TeXButton{\preprint}{\preprint{}}}%
%BeginExpansion
\preprint{}%
%EndExpansion
%

%TCIMACRO{\TeXButton{\title}{\title{Optical Potential Approach to $K^{+}%
%d$ Scattering at Low Energies}}}%
%BeginExpansion
\title{Optical Potential Approach to $K^{+}d$ Scattering at Low Energies}%
%EndExpansion
%Force line breaks with \\
%

%TCIMACRO{\TeXButton{\author}{\author{Takashi Takaki}}}%
%BeginExpansion
\author{Takashi Takaki}%
%EndExpansion
%

%TCIMACRO{\TeXButton{\email}{\email{takaki@onomichi-u.ac.jp}}}%
%BeginExpansion
\email{takaki@onomichi-u.ac.jp}%
%EndExpansion
%

%TCIMACRO{\TeXButton{\affiliation}{\affiliation{
%Department of Economics,Management and Information Science,
%Onomichi University, Hisayamada 1600, 722-0021 Onomichi, Japan
%}}}%
%BeginExpansion
\affiliation{
Department of Economics,Management and Information Science,
Onomichi University, Hisayamada 1600, 722-0021 Onomichi, Japan
}%
%EndExpansion
%

%TCIMACRO{\TeXButton{\date{\today}}{\date{\today}}}%
%BeginExpansion
%\date{\today}%
%EndExpansion
%It is always \today, today,
%but any date may be explicitly specified
%

%TCIMACRO{\TeXButton{%
%\begin{abstract}%
%}{\begin{abstract}}}%
%BeginExpansion
\begin{abstract}%
%EndExpansion

We study the $K^{+}d$ scattering at low energies using the optical potential.
Our optical potential consists of the first-order and second-order terms. The
total, integrated elastic and elastic differential cross sections at incident
kaon momenta below 800 MeV/c are calculated using our optical potential. We
found that our results are consistent with the Faddeev calculation as well as
the data and especially the second-order optical potential is essential to
reproduce them at low energies. We also discuss the multiple scattering
effects in this process.%

%TCIMACRO{\TeXButton{%
%\end{abstract}%
%}{\end{abstract}}}%
%BeginExpansion
\end{abstract}%
%EndExpansion
%

%TCIMACRO{\TeXButton{\pacs}{\pacs
%{11.80.-m; 13.75.Jz; 14.40.Df; 24.10.-i; 25.80.Nv}}}%
%BeginExpansion
%EndExpansion
%PACS, the Physics and Astronomy
%Classification Scheme.

%\keywords{Suggested keywords}%Use showkeys class option if keyword
%display desired
%

%TCIMACRO{\TeXButton{\maketitle}{\maketitle}}%
%BeginExpansion
\maketitle
%EndExpansion

\section{\textbf{Introduction}}

One of the important purposes for the studies of the $K^{+}d$ scattering is to
understand the $K^{+}N$ interaction, since the isospin zero $K^{+}N$ amplitude
or $K^{+}$-neutron amplitude can be obtained only through the $K^{+}d$
scattering. Recently some experimental studies\cite{ref1,ref2} have suggested
that the pentaquark resonance $\Theta(1540)$ with a narrow width might be
excited in this isospin channel, although its existence has not been confirmed
yet. As the $K^{+}$ meson is built from a $u\overline{s}$ quark state, it
cannot form the conventional three-quark resonance with a nucleon. So the
pentaquark resonance must have the exotic structure such as a $uudd\overline
{s}$ quark state and the coupling with the $K^{+}N$ system is expected to be
weak. In fact, the width of the $\Theta(1540)$ has been found to be less than
1 MeV\cite{ref3,ref4,ref5} through the analysis of$\ $the $K^{+}d$ reaction.
Therefore the $K^{+}N$ system is the weaker interacting system compared with
other meson-nucleon systems such as $\pi^{\pm}N$ and $K^{-}N$ which have the
strong couplings with resonant particles. The $K^{+}d$ scattering at low
energies is much simpler than the $\pi^{\pm}d$ and $K^{-}d$ scatterings since
the pion-absorption in the $\pi^{\pm}d$ scattering and the conversion to $\pi
N\Lambda$ and $\pi N\Sigma$ in the $K^{-}d$ scattering occur even at the
threshold. At incident kaon momenta below 600 MeV/c, where the pion production
does not occur, the $K^{+}d$ system has only three reactions, i.e., the
elastic scattering $K^{+}d\rightarrow K^{+}d$, the breakup reaction
$K^{+}d\rightarrow K^{+}pn$ and the charge exchange reaction $K^{+}%
d\rightarrow K^{0}pp$. Thus the analysis of the $K^{+}d$ scattering at low
energies is suited to rigorously examining the validity of various theoretical
models\cite{ref5,ref6,ref7,ref8}. One of them is the three-body calculation by
the Faddeev method\cite{ref7,ref8} where the multiple scattering effects have
been estimated accurately.

Because of the weak $K^{+}N$ interaction, the single scattering impulse
approximation is able to successfully describe the total cross sections at
incident momenta ($P_{lab})$ above $\sim$500 MeV/c which are obtained via the
optical theorem. Furthermore the elastic differential cross sections at the
low momentum transfer can be explained even at lower energies. However, the
single scattering impulse approximation explicitly violates the unitarity at
incident momenta below $\sim$200 MeV/c, where the integrated elastic cross
section is found to be larger than the total cross section\cite{ref7}, and
fails to describe the breakup differential cross section at the forward kaon
scattering angles\cite{ref8}. This means that the multiple scattering effects
should be incorporated in a theory to describe the $K^{+}d$ scattering
consistently, particularly, at low energies. If the pentaquark resonance
$\Theta(1540)$ would exist, one could see the signal in the cross sections of
the $K^{+}d$ reactions at $P_{lab}$ $\approx$450 MeV/c. In this energy region,
the multiple scattering effects are expected to still have a non-negligible
contribution\cite{ref4} and thus have to be considered in the study of the
$\Theta(1540)$ resonance.

In this paper we will investigate the $K^{+}d$ scattering by using the optical
potential defined in the multiple scattering theory of Watson\cite{ref9}. Our
optical potential consists of the first-order and second-order terms. The
second-order optical potential is constructed to include the multiple
scattering corrections such as the kaon rescattering, the Pauli correction and
the N-N interaction. One of the features of this approach is that it does not
violate the unitarity. Since this formulation has a simple structure,
furthermore, the calculation is more easily performed than the Faddeev
calculation. We will examine how our optical potential works for the $K^{+}d$
scattering and demonstrate the importance of the multiple scattering effects.
To do so, we calculate the energy dependence of the total cross sections,
integrated reaction cross sections and elastic cross sections and then compare
our calculations with the data. We show that our approach predicts the data
successfully as the Faddeev method and especially the second-order optical
potential is needed to describe both the data and the Faddeev calculation at
low energies.\ Indeed, the cross section calculated with only the first-order
optical potential is not consistent with the Faddeev calculation at low
energies in spite of the weak $K^{+}N$ interaction. Our present work is the
first application of the optical potential including the second-order term to
the $K^{+}d$ scattering. The optical potential up to second order is also
obtained from the Kerman-McManus-Thaler (KMT) \cite{ref10} theory. We will
discuss the relation between our approach and the KMT one.

Our paper is organized as follows: we present our formalism in Sec.2. In Sec.3
we show our calculations for the total, elastic and inelastic cross sections
and compare them with the data and the Faddeev calculation. In Sec.4 we
summarize our work.

\section{\textbf{Formalism}}

\subsection{Outline of the optical potential approach}

To calculate the cross sections of the $K^{+}d$ scattering, we use the optical
potential derived from the multiple scattering theory. There are two
formulations for the multiple scattering theory, i.e., the Watson\cite{ref9}
and the KMT\cite{ref10} theories, which give the identical transition
amplitude if any approximations are not used. In this work, we will employ the
Watson theory because the physical interpretation of the optical potential is
clear. Our optical potential will be constructed to incorporate the multiple
scattering corrections such as the rescattering, the Pauli effect and the N-N
interaction. To do so, the second-order optical potential is considered in
addition to the usual first-order optical potential. Now we start from making
a brief review of the Watson formulation.

For the scattering of a positive kaon from A identical nucleons, the
transition amplitude $T$ is a solution of the Lippmann-Schwinger equation%
\begin{equation}
T=%
%TCIMACRO{\dsum \limits_{i=1}^{A}}%
%BeginExpansion
{\displaystyle\sum\limits_{i=1}^{A}}
%EndExpansion
v_{i}+%
%TCIMACRO{\dsum \limits_{i=1}^{A}}%
%BeginExpansion
{\displaystyle\sum\limits_{i=1}^{A}}
%EndExpansion
v_{i}\frac{\mathcal{A}}{e}T, \label{2.1}%
\end{equation}
where%
\begin{equation}
e=E-K_{0}-H_{A}+i\epsilon. \label{2.2}%
\end{equation}
Here $E$ is the total energy, $K_{0}$ is the kaon kinetic energy and $H_{A}$
is the Hamiltonian of the target nucleus. The two-body potential $v_{i}$
describes the interaction between the kaon and \textit{i}th nucleon.
$\mathcal{A\ }$is a projection operator onto the antisymmetric subspace of the
Hilbert space. As far as one works in the antisymmetric subspace, the operator
$\mathcal{A}$\textit{\ }is not necessary in Eq.(\ref{2.1}) since $T$ and
$\sum_{i=1}^{A}v_{i}$\textit{\ }are symmetric operators. For later
convenience, however, it is inserted explicitly in Eq.(\ref{2.1}).

Now we introduce a projection operator $P$ which projects onto the nuclear
ground state and $Q$ is defined by%
\begin{equation}
Q=\mathcal{A}-P. \label{2.3}%
\end{equation}
By using these operators, Eq.(\ref{2.1}) is rewritten in terms of the optical
potential $U$ as%
\begin{equation}
T=U+U\frac{P}{e}T, \label{2.4}%
\end{equation}
where $U$ is given by%
\begin{equation}
U=\sum_{i=1}^{A}U_{i}, \label{2.5}%
\end{equation}
with
\begin{equation}
U_{i}=\widehat{t_{i}}+\widehat{t_{i}}\frac{Q}{e}\sum_{j\neq i}U_{j}.
\label{2.6}%
\end{equation}
The kaon-nucleon T matrix $\widehat{t_{i}}$ in Eq.(\ref{2.6}) is defined by%
\begin{equation}
\widehat{t_{i}}=v_{i}+v_{i}\frac{Q}{e}\widehat{t_{i}}. \label{2.7}%
\end{equation}

To evaluate the optical potential $U$, we define another kaon-nucleon T matrix
$t_{i}$ such that%
\begin{equation}
t_{i}=v_{i}+v_{i}\frac{1}{e}t_{i}, \label{2.8}%
\end{equation}
where the kaon propagates in the space of both antisymmetric and
non-antisymmetric states of the nucleus. The relation between many-body
operators $\widehat{t_{i}}$ and $t_{i}$ is%
\begin{equation}
\widehat{t_{i}}=t_{i}-t_{i}\frac{1-\mathcal{A}+P}{e}\widehat{t_{i}}.
\label{2.9}%
\end{equation}
Here the operator $1-\mathcal{A}$ projects onto the Pauli-violating states.
The second term of Eq.(\ref{2.9}) appears to remove the contribution of the
transition to the Pauli-violating states and the ground state from $t_{i}$. If
the term proportional to $P$ is neglected, the coherent rescattering is
overcounted in the calculation of Eq.(\ref{2.4}).

Before constructing our model, let us discuss the relation between the Watson
and KMT formulations. In the KMT theory (see Appendix), the following
many-body operator $\tau_{i}$ is used to define the optical potential:%

\begin{equation}
\tau_{i}=t_{i}-t_{i}\frac{1-\mathcal{A}}{e}\tau_{i}, \label{2.9-1}%
\end{equation}
where the projection operator $P$ does not appear since the coherent
rescattering is counted in a different way. The relation between the operators
$\widehat{t_{i}}$ and $\tau_{i}$ is given by
\begin{equation}
\widehat{t_{i}}=\tau_{i}-\tau_{i}\frac{P}{e}\widehat{t_{i}}. \label{2.9-2}%
\end{equation}
The first-order optical potentials are given by $\sum_{i=1}^{A}\widehat{t_{i}%
}$ for the Watson approach and $\frac{A-1}{A}\sum_{i=1}^{A}\tau_{i}$ for the
KMT approach, respectively. Within this first-order expansion, the two
approaches give the identical transition amplitude as pointed out in
Ref.\cite{ref11}.\ Even if further approximations are assumed, this is still
correct as far as the relation (\ref{2.9-2}) holds. In the actual
calculations, however, one usually uses the impulse approximation where
$\widehat{t_{i}}$ and $\tau_{i}$ are replaced by the free two-body T matrix
$t_{i}^{free}$. In this case, the relation (\ref{2.9-2}) does not hold and
therefore the two approaches give different transition amplitudes. In fact, it
has been shown from the studies\cite{ref12,ref13} of the pion-deuteron
scattering that the KMT first-order optical potential is superior to the
Watson one. So it is necessary to go beyond the first-order optical potential
in the impulse approximation and take into account the second term of
Eq.(\ref{2.9}) in order to get the reliable results within the Watson
formulation. There are several models taking account of this term. In the
delta-hole model\cite{ref14}, the second term of Eq.(\ref{2.9}) is
incorporated by adding a Fock term to the delta-hole propagator. In the model
of Ref.\cite{ref11}, the channel-coupled equations are derived from
Eq.(\ref{2.9}) and are solved to get the first-order optical potential. In our
work, we will expand the T matrix $\widehat{t_{i}}$ of Eq.(\ref{2.9}) in terms
of $t_{i}$ and consider it to second order.

Now we construct the optical potential which will be used in our calculations.
By substituting Eq.(\ref{2.9}) into Eq.(\ref{2.6}), the optical potential $U$
can be given in terms of $t_{i}$ by%
\begin{align}
U  &  =\sum_{i}t_{i}-\sum_{i}t_{i}\frac{1-\mathcal{A}+P}{e}t_{i}+\sum_{i\neq
j}t_{i}\frac{Q}{e}t_{j}+\cdots\label{2.10}\\
&  =\sum_{i}t_{i}+\sum_{i\neq j}t_{i}\frac{1}{e}t_{j}-\sum_{i,j}t_{i}\frac
{P}{e}t_{j}+\cdots, \label{2.11}%
\end{align}
which is explicitly written up to second order in $t_{i}$. We note that the
term including the operator $1-\mathcal{A}$ does not appear in the
second-order term of Eq.(\ref{2.11}), because the matrix element of
$\sum_{i,j}t_{i}\frac{1-\mathcal{A}}{e}t_{j}$ between the nuclear
antisymmetric states vanishes.

We now consider the deuteron as the target nucleus and introduce additional
approximations to derive the optical potential in our approach. In the impulse
approximation, $t_{i}$ ($i=1,2$) is replaced by the free two-body T matrix
$t_{i}^{free}$ defined by%
\begin{equation}
t_{i}^{free}=v_{i}+v_{i}\frac{1}{e_{0}}t_{i}^{free}, \label{2.12}%
\end{equation}
with%
\begin{equation}
e_{0}=E-K_{0}-K_{1}-K_{2}+i\epsilon, \label{2.13}%
\end{equation}
where $K_{i}$ ($i=1,2$) is the kinetic energy of \textit{i}th nucleon and
$E-K_{j}(j\neq i)$ is the collision energy for the kaon-\textit{i}th nucleon
subsystem. In our approach, the effect of the nucleon-nucleon (N-N)
interaction is taken into account because it is important at low energies and
particularly it leads to the non-negligible final state interaction in the
breakup reaction. Since $H_{A}$ in Eq.(\ref{2.2}) is equal to $K_{1}%
+K_{2}+v_{12}$ where $v_{12}$ is the N-N interaction, the many-body Green
function $1/e$ can be expressed in terms of the two-nucleon T matrix $t_{NN}$
as%
\begin{equation}
\frac{1}{e}=\frac{1}{e_{0}}+\frac{1}{e_{0}}t_{NN}\frac{1}{e_{0}}, \label{2.14}%
\end{equation}
with%
\begin{equation}
t_{NN}=v_{12}+v_{12}\frac{1}{e_{0}}t_{NN}, \label{2.15}%
\end{equation}
where the collision energy of the two-nucleon subsystem is $E-K_{0}$. Thus the
many-body operator $t_{i}$ is expressed as%
\begin{equation}
t_{i}=t_{i}^{free}+t_{i}^{free}\frac{1}{e_{0}}t_{NN}\frac{1}{e_{0}}t_{i}.
\label{2.16}%
\end{equation}
Substitution of Eqs.(\ref{2.14}) and (\ref{2.16}) into Eq.(\ref{2.11}) yields%
\begin{equation}
U=U^{(1)}+U^{(2)}+\cdots, \label{2.17}%
\end{equation}
where%
\begin{align}
U^{(1)}  &  =\sum_{i=1}^{2}t_{i}^{free},\label{2.18}\\
U^{(2)}  &  =\sum_{i\neq j}t_{i}^{free}\frac{1}{e_{0}}t_{j}^{free}+\sum
_{i,j}t_{i}^{free}(\frac{1}{e_{0}}t_{NN}\frac{1}{e_{0}}-\frac{P}{e}%
)t_{j}^{free}. \label{2.19}%
\end{align}
Here $U^{(1)}$ and $U^{(2)}$ are the first-order and second-order optical
potentials, respectively. The higher-order potentials are neglected in our
calculation. The second-order potential consists of three terms as%
\begin{equation}
U^{(2)}=U_{d}^{(2)}+U_{n}^{(2)}-U_{c}^{(2)}, \label{2.20}%
\end{equation}
where $U_{d}^{(2)}$ describes the double scattering, and $U_{n}^{(2)}$ and
$U_{c}^{(2)}$ are the N-N scattering and the coherent rescattering terms,
respectively. The quantity $U_{n}^{(2)}-U_{c}^{(2)}$ represents the effect of
the modified N-N scattering where some contribution of the N-N bound state is
excluded. We notice that in the vicinity of the N-N bound state pole at
$E_{N}=-E_{B}$, where $E_{N}=E-K_{0}-K_{CM}$ and $K_{CM}$ is the kinetic
energy of the center of mass of two nucleons, we may write,%
\begin{equation}
\frac{1}{e_{0}}t_{NN}\frac{1}{e_{0}}\cong\frac{P}{e}. \label{2.19-1}%
\end{equation}

Here we consider the relation between the Watson and KMT transition
amplitudes, i.e., $T$ and $T_{KMT}$. Let us assume that the corresponding
optical potentials are $U\cong U^{(1)}+U^{(2)}$ and $U_{KMT}\cong
U_{KMT}^{(1)}+U_{KMT}^{(2)}$ (see Appendix), respectively. These transition
amplitudes are expanded in powers of $U$ or $U_{KMT}$ as
\begin{align}
T  &  =U+U\frac{P}{e}U+\cdots,\label{2.20-1}\\
T_{KMT}  &  =2(U_{KMT}+U_{KMT}\frac{P}{e}U_{KMT}+\cdots). \label{2.20-2}%
\end{align}
Using the above equations, the amplitudes $T$ and $T_{KMT}$ can be rewritten
in powers of \ $t_{i}^{free}$. Then one finds the following relation,%
\begin{equation}
T=T_{KMT}+O((t^{free})^{3}). \label{2.20-3}%
\end{equation}
$T$ is equal to $T_{KMT}$ up to second order in $t^{free}$. If the second term
on the right side of Eq.(\ref{2.20-3}) would be small, the two approaches
would give approximately the identical transition amplitude. We have
numerically checked that this is true for the $K^{+}d$ scattering. We note
that $T\cong T_{KMT}\cong U^{(1)}$ at kaon momenta above $\sim$500 MeV/c where
the single scattering impulse approximation works well.

The purpose of this work is to evaluate the total and integrated elastic cross
sections of the $K^{+}d$ scattering. These are obtained by solving the
equation (\ref{2.4}) where the optical potential $U$ is given by
Eqs.(\ref{2.18}) and (\ref{2.19}). The total cross section is calculated via
the optical theorem as%
\begin{equation}
\sigma_{tot}=\frac{4\pi}{k}\operatorname{Im}f_{Kd}(\mathbf{k},\mathbf{k}),
\label{2.21}%
\end{equation}
where the scattering amplitude $f_{Kd}$ is given by
\begin{equation}
f_{Kd}(\mathbf{k}^{\prime},\mathbf{k})=-\frac{2\omega E_{d}}{4\pi
W}\left\langle \mathbf{k}^{\prime}d\right\vert T\left\vert \mathbf{k}%
d\right\rangle . \label{2.22}%
\end{equation}
Here $\mathbf{k(k}^{\prime})$ is the initial (final) kaon momentum, $\omega$
and $E_{d}$ are the total energies of the kaon and the deuteron in the
kaon-deuteron center of mass (c.m.) frame, and $W(=$ $\omega+E_{d})$ is the
total energy of the kaon-deuteron system. The spin quantum numbers are
implicitly included. The integrated elastic cross section for the unpolarized
deuteron is obtained by integrating the differential cross section over the
angle as%
\begin{align}
\sigma_{el}  &  =\int\frac{d\sigma}{d\Omega}d\Omega\nonumber\\
&  =\int(\frac{1}{3}\sum_{spin}\left\vert f_{Kd}(\mathbf{k}^{\prime
},\mathbf{k})\right\vert ^{2})d\Omega, \label{2.23}%
\end{align}
where $\left\vert f_{Kd}\right\vert ^{2}$ is summed over the initial and final
spin orientations.

\subsection{The method of calculation}

Our numerical calculations will be performed in the momentum space
representation. The Lippmann-Schwinger equation (\ref{2.4}) in the
kaon-deuteron center of mass frame is expressed as \begin{widetext}%
\begin{align}
\left\langle \mathbf{k}^{\prime}d\left\vert T\right\vert \mathbf{k}%
d\right\rangle  &  =\left\langle \mathbf{k}^{\prime}d\left\vert
U\right\vert \mathbf{k}d\right\rangle +%
%TCIMACRO{\dint }%
%BeginExpansion
{\displaystyle\int}
%EndExpansion
\frac{d^{3}k^{\prime\prime}}{(2\pi)^{3}}\left\langle \mathbf{k}^{\prime
}d\left\vert U\right\vert \mathbf{k}^{\prime\prime}d\right\rangle
G_{p}(W,\mathbf{k}^{\prime\prime})\left\langle \mathbf{k}^{\prime\prime
}d\left\vert T\right\vert \mathbf{k}d\right\rangle ,\label{2.24}\\
G_{p}(W,\mathbf{k}^{\prime\prime}) &  =\frac{1}{W-\omega^{\prime\prime}%
-E_{d}(k^{\prime\prime})+i\epsilon}.\label{2.25}%
\end{align}
\end{widetext}Here the spin and isospin are omitted for simplicity. This
equation is solved by decomposing into the partial waves. To calculate the
optical potential $\left\langle \mathbf{k}^{\prime}d\left\vert U\right\vert
\mathbf{k}d\right\rangle $ given by Eqs.(\ref{2.18}) and (\ref{2.19}), one
needs the off-shell kaon-nucleon T matrix $t^{free}$ and the off-shell
nucleon-nucleon T matrix $t_{NN}$. They are taken to be of separable form.

The kaon-nucleon T matrix in the general frame is%
\begin{align}
\left\langle \mathbf{k}^{\prime}\mathbf{p}^{\prime}\left\vert t^{free}%
\right\vert \mathbf{kp}\right\rangle  &  =\frac{m}{\sqrt{4\omega^{\prime
}\omega E_{p^{\prime}}E_{p}}}\mathcal{M}(s,\mathbf{\kappa}^{\prime
},\mathbf{\kappa}),\label{2.26}\\
\mathcal{M}(s,\mathbf{\kappa}^{\prime},\mathbf{\kappa})  &  =-\frac{4\pi
\sqrt{s}}{m}f_{KN}(s,\mathbf{\kappa}^{\prime},\mathbf{\kappa}), \label{2.26-1}%
\end{align}
where $f_{KN}$ is the scattering amplitude consisting of the non spin-flip and
spin-flip terms, and $\mathbf{k}$, $\mathbf{k}^{\prime}$ ,$\mathbf{p}$ and
$\mathbf{p}^{\prime}$ are the momenta of the initial kaon, the final kaon, the
initial nucleon and the final nucleon, respectively. Furthermore, $\omega
$,$\omega^{\prime}$,$E_{p^{\prime}}$ and $E_{p}$ are the energies of the
initial kaon, the final kaon, the initial nucleon and the final nucleon,
respectively, and $m$ and $s$ are the nucleon mass and the invariant mass
squared of the kaon-nucleon system and $\mathbf{\kappa}$ and $\mathbf{\kappa
}^{\prime}$ are the momenta of the initial and final kaons in the kaon-nucleon
c.m. frame. The partial wave amplitude $\mathcal{M}_{lj}$ is given as%
\begin{align}
\mathcal{M}_{lj}(s,\kappa^{\prime},\kappa)  &  =g_{l}(\kappa^{\prime}%
)G_{lj}(s)g_{l}(\kappa),\label{2.27}\\
G_{lj}(s)  &  =-\frac{4\pi\sqrt{s}}{m}\frac{f_{lj}(s)}{(g_{l}(\kappa_{0}%
))^{2}},\label{2.28}\\
f_{lj}(s)  &  =\frac{e^{2i\delta_{lj}(s)}-1}{2i\kappa_{0}}, \label{2.28-1}%
\end{align}
where $l$ and $j$ are the orbital and total angular momentum, $\kappa_{0}$ is
the on-shell momentum evaluated from $s$ and $\delta_{lj}(s)$ is the phase
shift. For the kaon-deuteron scattering, the quantity $s$ in Eq.(\ref{2.27})
is taken as%
\begin{equation}
s=(W-E_{p_{R}})^{2}-p_{R}^{2}, \label{2.29}%
\end{equation}
where $p_{R}$ and $E_{p_{R}}$ is the momentum and energy of the spectator
nucleon. Here the spectator nucleon is assumed to be on-shell. The form of
Eq.(\ref{2.28}) is used for the physical region $s\geqq(m_{K}+m)^{2}$ where
$m_{K}$ is the kaon mass. For the unphysical region $s<(m_{K}+M)^{2}$, on the
other hand, $G_{lj}$ is assumed as%
\begin{equation}
G_{lj}(s)=G_{lj}((m_{K}+m)^{2})\frac{1}{(2-s/(m_{K}+m)^{2})^{2}}. \label{2.30}%
\end{equation}
The form factor $g_{l}$ is taken as
\begin{equation}
g_{l}(\kappa)=\frac{\kappa^{l}}{(\beta^{2}+\kappa^{2})^{n}}, \label{2.31}%
\end{equation}
with $\beta=1$ GeV/c. Here $n=1$ is used for $l=0$ and $1$, and $n=2$ is used
for $l=2$ and $3$. We employ the same prescription used by
Garcilazo\cite{ref7}.

For the nucleon-nucleon T matrix, we use the separable form made by means of
the Ernst-Shakin-Thaler (EST) method\cite{ref15}. Since we are mainly
interested in the total cross section and integrated elastic cross section at
low energies, we will take into account only the S-wave contribution in the
nucleon-nucleon interaction, and disregard the coupling to the D-wave. In this
approximation, the nucleon-nucleon T matrix is given as%
\begin{align}
\left\langle p^{\prime}\left\vert t_{NN}(E_{N})\right\vert p\right\rangle  &
=g_{N}(p^{\prime})\tau_{NN}(E_{N})g_{N}(p),\label{2.32}\\
\lbrack\tau_{NN}(E_{N})]^{-1}  &  =-1+\frac{\mu}{\pi^{2}}%
%TCIMACRO{\dint \nolimits_{0}^{\infty}}%
%BeginExpansion
{\displaystyle\int\nolimits_{0}^{\infty}}
%EndExpansion
\frac{q^{2}(g_{N}(q))^{2}}{q^{2}-k_{N}^{2}-i\varepsilon}dq, \label{2.33}%
\end{align}
where $\mu=m/2$ , $k_{N}^{2}=2\mu E_{N}$ and $p(p^{\prime})$ is the initial
(final) relative momentum of the nucleon-nucleon system. The spin and isospin
are again omitted. Here the form factor $g_{N}$ is defined by $g_{N}%
(p)=\sqrt{2\pi^{2}}g(p)$ in which $g$ is the analytical form factor obtained
by using the Paris nucleon-nucleon potential\cite{ref16}. In this method, the
form factor for the $^{3}S_{1}$ channel is related to the deuteron wave
function as%
\begin{equation}
\psi_{d}(\mathbf{p)}=\frac{2\mu Ng_{N}(p)}{k_{B}^{2}+p^{2}}, \label{2.34}%
\end{equation}
where $k_{B}^{2}=2\mu E_{B}$. Here $E_{B}$ is the binding energy of the
deuteron and $N$ is a normalization constant.

Now we discuss how the optical potential is evaluated. The first-order optical
potential in the kaon-deuteron c.m. frame is written as \begin{widetext}%
\begin{align}
\left\langle \mathbf{k}^{\prime}d\left\vert U^{(1)}\right\vert
\mathbf{k}d\right\rangle  &  =\left\langle \mathbf{k}^{\prime}d\left\vert
\sum_{i=1}^{2}t_{i}^{free}\right\vert \mathbf{k}d\right\rangle \nonumber\\
&  =%
%TCIMACRO{\dint }%
%BeginExpansion
{\displaystyle\int}
%EndExpansion
\frac{d^{3}\kappa_{d}}{(2\pi)^{3}}\psi_{d}(\mathbf{\kappa}_{d}^{\prime
}\mathbf{)}\left\langle \mathbf{k}^{\prime}\mathbf{p}_{1}^{\prime}\left\vert
(t_{p}^{free}(s_{1})+t_{n}^{free}(s_{1}))\right\vert \mathbf{kp}%
_{1}\right\rangle \psi_{d}(\mathbf{\kappa}_{d}\mathbf{),}\label{2.36}%
\end{align}
\end{widetext}where $\mathbf{\kappa}_{d}(\mathbf{\kappa}_{d}^{\prime})$ is the
relative momentum of the two-nucleon, $\mathbf{p}_{1}\mathbf{(p}_{1}^{\prime
})$ is the momentum of the struck nucleon and the spin of the deuteron is
omitted for simplicity. The free kaon-nucleon T matrices $t_{p}^{free}$ and
$t_{n}^{free}$ describe the processes $K^{+}p\longrightarrow K^{+}p$ and
$K^{+}n\longrightarrow K^{+}n$, respectively and are evaluated at the
invariant mass squared $s_{1}$. Here $s_{1}=(W-E_{p_{2}})^{2}-p_{2}^{2}$ and
$p_{2}$ is the momentum of the spectator nucleon. The momenta appeared in the
integrand are defined in a non-relativistic way. All of them are given in
terms of $\mathbf{\kappa}_{d},$ $\mathbf{k}$ and $\mathbf{k}^{\prime}$ if the
momentum conservation law is used. The numerical calculation of Eq.(\ref{2.36}%
) is performed without any factorization. In the energy region we study, the
non spin-flip contributions dominate over the spin-flip contributions in the
kaon-nucleon T matrix. Taking account of this fact, the first-order potential
with only the S-wave and the non spin-flip P-wave term is used to solve the
equation (\ref{2.24}), while for the spin-flip P-wave term and the D-wave
term, the single scattering impulse approximation is used. For the on-shell
kaon-nucleon scattering amplitude, we use the phase shifts of
Martin\cite{ref17} or Hyslop \textit{et al.}\cite{ref18}. We note that the T
matrix $t_{p}^{free}(s_{1})+t_{n}^{free}(s_{1})$ in Eq.(\ref{2.36}) is
expressed as $1/2t_{I=0}^{free}(s_{1})+3/2t_{I=1}^{free}(s_{1})$\ where
$t_{I=0}^{free}$ and $t_{I=1}^{free}$ are $I=0$ and $I=1$ isospin T matrices,
respectively. This relation indicates that the $I=1$ $KN$ interaction
plays\ more important role than the $I=0$ $KN$ interaction in the $K^{+}d$
elastic scattering.

The second-order optical potential is constructed with the S-wave kaon-nucleon
interaction and the S-wave nucleon-nucleon interaction, since these
contributions are expected to be dominant at low energies. In this
approximation, thus, the second-order potential is spin-independent. So the
spin of the deuteron is suppressed in the following expressions. The double
scattering term $U_{d}^{(2)}$ is written as \begin{widetext}%
\begin{align}
\left\langle \mathbf{k}^{\prime}d\left\vert U_{d}^{(2)}\right\vert
\mathbf{k}d\right\rangle  &  =\left\langle \mathbf{k}^{\prime}d\left\vert
\sum_{i\neq j}t_{i}^{free}\frac{1}{e_{0}}t_{j}^{free}\right\vert
\mathbf{k}d\right\rangle \nonumber\\
&  =\int\frac{d^{3}k^{\prime\prime}}{(2\pi)^{3}}\frac{d^{3}p_{r}}{(2\pi)^{3}%
}\psi_{d}(\mathbf{\kappa}_{d}^{\prime}\mathbf{)}G_{0}(k^{\prime\prime}%
,p_{r},k)\{\left\langle \mathbf{k}^{\prime}\mathbf{p}_{2}^{\prime}\left\vert
t_{p}^{free}(s_{2})\right\vert \mathbf{k}^{\prime\prime}\mathbf{p}%
_{2}\right\rangle \left\langle \mathbf{k}^{\prime\prime}\mathbf{p}_{1}%
^{\prime}\left\vert t_{n}^{free}(s_{1})\right\vert \mathbf{kp}_{1}%
\right\rangle \nonumber\\
&  +\left\langle \mathbf{k}^{\prime}\mathbf{p}_{2}^{\prime}\left\vert
t_{n}^{free}(s_{2})\right\vert \mathbf{k}^{\prime\prime}\mathbf{p}%
_{2}\right\rangle \left\langle \mathbf{k}^{\prime\prime}\mathbf{p}_{1}%
^{\prime}\left\vert t_{p}^{free}(s_{1})\right\vert \mathbf{kp}_{1}%
\right\rangle \nonumber\\
&  -\left\langle \mathbf{k}^{\prime}\mathbf{p}_{2}^{\prime}\left\vert
t_{ex}^{free}(s_{2})\right\vert \mathbf{k}^{\prime\prime}\mathbf{p}%
_{2}\right\rangle \left\langle \mathbf{k}^{\prime\prime}\mathbf{p}_{1}%
^{\prime}\left\vert t_{ex}^{free}(s_{1})\right\vert \mathbf{kp}_{1}%
\right\rangle \}\psi_{d}(\mathbf{\kappa}_{d}\mathbf{),}\label{2.37}%
\end{align}
where%
\begin{equation}
G_{0}^{-1}(k^{\prime\prime},p_{r},k)=\omega+\frac{k^{2}}{4m}-E_{B}%
-\omega^{\prime\prime}-\frac{k^{\prime\prime2}}{4m}-\frac{p_{r}^{2}}%
{m}+i\varepsilon.\label{2.38}%
\end{equation}
Here $p_{r}$ is the relative momentum of two-nucleon, and $s_{1}=(W-E_{p_{2}%
})^{2}-p_{2}^{2}$ \ and $s_{2}=(W-E_{p_{1}^{\prime}})^{2}-p_{1}^{\prime2}$
where $p_{2}$ and $p_{1}^{\prime}$ are the momenta of the spectator nucleon.
The free kaon-nucleon T matrix $t_{ex}^{free}$ describes the charge exchange
process $K^{+}n\longrightarrow K^{0}p$ ( $K^{0}p\longrightarrow K^{+}n$). The
N-N scattering term $U_{n}^{(2)}$ is written as%
\begin{align}
\left\langle \mathbf{k}^{\prime}d\left\vert U_{n}^{(2)}\right\vert
\mathbf{k}d\right\rangle  &  =\left\langle \mathbf{k}^{\prime}d\left\vert
\sum_{i,j}t_{i}^{free}\frac{\mathcal{A}}{e_{0}}t_{NN}\frac{\mathcal{A}}{e_{0}%
}t_{j}^{free}\right\vert \mathbf{k}d\right\rangle \nonumber\\
&  =\int\frac{d^{3}k^{\prime\prime}}{(2\pi)^{3}}\tau_{NN}(E_{N}%
(k^{\prime\prime}))\nonumber\\
&  \times\int\frac{d^{3}p_{r}^{\prime}}{(2\pi)^{3}}\psi_{d}(\mathbf{\kappa
}_{d}^{\prime}\mathbf{)}\left\langle \mathbf{k}^{\prime}\mathbf{p}_{2}%
^{\prime}\left\vert (t_{p}^{free}(s_{2})+t_{n}^{free}(s_{2}))\right\vert
\mathbf{k}^{\prime\prime}\mathbf{p}_{2}^{\prime\prime}\right\rangle
G_{0}(k^{\prime\prime},p_{r}^{\prime},k)g_{N}(p_{r}^{\prime})\nonumber\\
&  \times\int\frac{d^{3}p_{r}}{(2\pi)^{3}}g_{N}(p_{r})G_{0}(k^{\prime\prime
},p_{r},k)\left\langle \mathbf{k}^{\prime\prime}\mathbf{p}_{1}^{\prime\prime
}\left\vert (t_{p}^{free}(s_{1})+t_{n}^{free}(s_{1}))\right\vert
\mathbf{kp}_{1}\right\rangle \psi_{d}(\mathbf{\kappa}_{d}\mathbf{),}%
\label{2.39}%
\end{align}
where
\begin{equation}
E_{N}(k^{\prime\prime})=\omega+\frac{k^{2}}{4m}-E_{B}-\omega^{\prime\prime
}-\frac{k^{\prime\prime2}}{4m},\label{2.39-1}%
\end{equation}
and the coherent rescattering term $U_{c}^{(2)}$ is
\begin{align}
\left\langle \mathbf{k}^{\prime}d\left\vert U_{c}^{(2)}\right\vert
\mathbf{k}d\right\rangle  &  =\left\langle \mathbf{k}^{\prime}d\left\vert
\sum_{i,j}t_{i}^{free}\frac{P}{e}t_{j}^{free}\right\vert \mathbf{k}%
d\right\rangle \nonumber\\
&  =\int\frac{d^{3}k^{\prime\prime}}{(2\pi)^{3}}G_{c}(k^{\prime\prime
},k)\nonumber\\
&  \times\int\frac{d^{3}p_{r}^{\prime}}{(2\pi)^{3}}\psi_{d}(\mathbf{\kappa
}_{d}^{\prime}\mathbf{)}\left\langle \mathbf{k}^{\prime}\mathbf{p}_{2}%
^{\prime}\left\vert (t_{p}^{free}(s_{2})+t_{n}^{free}(s_{2}))\right\vert
\mathbf{k}^{\prime\prime}\mathbf{p}_{2}^{\prime\prime}\right\rangle \psi
_{d}(p_{r}^{\prime})\nonumber\\
&  \times\int\frac{d^{3}p_{r}}{(2\pi)^{3}}\psi_{d}(p_{r})\left\langle
\mathbf{k}^{\prime\prime}\mathbf{p}_{1}^{\prime\prime}\left\vert (t_{p}%
^{free}(s_{1})+t_{n}^{free}(s_{1}))\right\vert \mathbf{kp}_{1}\right\rangle
\psi_{d}(\mathbf{\kappa}_{d}\mathbf{),}\label{2.40}%
\end{align}
where%
\begin{equation}
G_{c}^{-1}(k^{\prime\prime},k)=\omega+\frac{k^{2}}{4m}-\omega^{\prime\prime
}-\frac{k^{\prime\prime2}}{4m}+i\varepsilon.\label{2.41}%
\end{equation}
\end{widetext}

\section{\textbf{Results}}

In this section, we will discuss the results calculated with the optical
potential consisting of the first-order and second-order terms. The optical
potential is evaluated by taking into account a three-body kinematics fully
and without any factorization in the momentum integration. To treat the
singularity properly, the usual subtraction procedure is used. As the
second-order potential is expected to be important only at low energies, we
take into account only the S-wave kaon-nucleon and nucleon-nucleon
interactions for it. As the S-wave kaon-nucleon interaction is described by
the non spin-flip amplitude, only the N-N interaction for the $^{3}S_{1}$
channel is included in this approximation. Consequently the N-N D-wave
contribution in the wave functions for the continuum and bound states is
ignored. This approximation is appropriate to describe the total and
integrated $K^{+}d$ elastic cross sections, since the scattering amplitude
$f_{Kd}$ at the low-momentum transfer dominantly contributes to them.%
%TCIMACRO{\FRAME{ftbpFU}{3.039in}{2.2295in}{0pt}{\Qcb{Total and integrated
%elastic $K^{+}d$ cross sections. The solid lines are the results of the full
%calculation. The dashed and dash-dotted lines are the results of the
%first-order optical potential and the single scattering impulse approximation,
%respectively and the dotted line is the result of Eq.(\ref{3.2}). Phase shifts
%of Martin\cite{ref17} are used in these calculations. The data of total cross
%sections are from Refs.\cite{ref8,ref19,ref20,ref21,ref22,ref23,ref24}%
%(circles). The data of integrated elastic cross sections are from
%Refs.\cite{ref8,ref25,ref26}(triangles). }}{\Qlb{fig-1}}{fig-1.eps}%
%{\special{ language "Scientific Word";  type "GRAPHIC";
%maintain-aspect-ratio TRUE;  display "USEDEF";  valid_file "F";
%width 3.039in;  height 2.2295in;  depth 0pt;  original-width 9.3763in;
%original-height 6.8554in;  cropleft "0";  croptop "1";  cropright "1";
%cropbottom "0";  filename 'figure/fig-1.eps';file-properties "XNPEU";}}}%
%BeginExpansion
\begin{figure}
[ptb]
\begin{center}
\includegraphics[
height=2.2295in,
width=3.039in
]%
{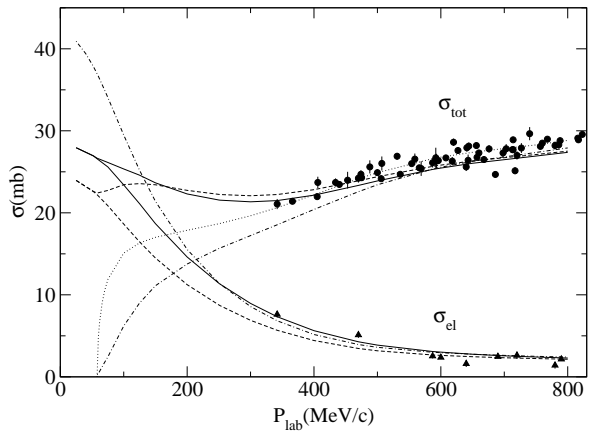}%
\caption{Total and integrated elastic $K^{+}d$ cross sections. The solid lines
are the results of the full calculation. The dashed and dash-dotted lines are
the results of the first-order optical potential and the single scattering
impulse approximation, respectively and the dotted line is the result of
Eq.(\ref{3.2}). Phase shifts of Martin\cite{ref17} are used in these
calculations. The data of total cross sections are from
Refs.\cite{ref8,ref19,ref20,ref21,ref22,ref23,ref24}(circles). The data of
integrated elastic cross sections are from Refs.\cite{ref8,ref25,ref26}%
(triangles). }%
\label{fig-1}%
\end{center}
\end{figure}
%EndExpansion

Now we show the calculations of the total and integrated elastic cross
sections for the $K^{+}d$ scattering with the experimental data in Fig.1. The
cross sections are plotted at incident momenta up to 800 MeV/c. In our
calculations, the coulomb interaction is not included and the $K^{+}N$ \ phase
shifts of Martin\cite{ref17} are used. \ The solid lines correspond to the
full calculations including the first- and second-order optical potentials.
The agreement with the data is satisfactory for both the total and elastic
cross sections. For the total cross section, the theory seems to be small
compared with several data at higher energies, but one can not say that there
is a discrepancy between the theory and the data since the data are scattered.
The dashed lines are the calculations including only the first-order optical
potential. The difference between these two lines shows the size of the
second-order potential effect. This effect is most important at lower
energies, especially near the threshold. For the total cross section, we find
that the second-order potential effect decreases its magnitude slightly at
$P_{lab}\gtrapprox$ 150 MeV/c, but it increases at $P_{lab}$ $\lessapprox$
150MeV/c. For the integrated elastic cross section, this effect increases it
and such tendency becomes stronger as the energy is lower. Generally speaking,
the second-order potential effect is important for the elastic cross section
than the total one but it is small at $P_{lab}\gtrapprox$ 500MeV/c for both
cases. The dash-dotted lines correspond to the calculations of the single
scattering impulse approximation where the transition amplitude $T_{IA}$ is
given as%
\begin{equation}
T_{IA}=\left\langle \mathbf{k}^{\prime}d\left\vert U^{(1)}\right\vert
\mathbf{k}d\right\rangle . \label{3.1}%
\end{equation}
In this approximation, the unitarity is explicitly violated at low momenta
where the elastic cross section is larger than the total cross section as
shown in Fig.1. From the comparison between the dash-dotted and dashed lines,
we find that the coherent rescattering effect has a significant contribution
for both the total and elastic cross sections at momenta below $\sim$500 MeV/c
and drastically changes the size of cross sections at low energies so as to
recover the unitarity. We consider further approximation to the total cross
section. We calculate it by factorizing the kaon-nucleon T matrix
$t_{i}^{free}$out of integral of Eq.(\ref{2.36}). Here the T matrix is
evaluated at $\mathbf{p}_{1}=-\mathbf{k}/2$ and its form factor is taken to be
$g_{l}=1$. Using the optical theorem, one gets%
\begin{equation}
\sigma_{K^{+}d}^{tot}=K(\sigma_{K^{+}p}^{tot}+\sigma_{K^{+}n}^{tot}),
\label{3.2}%
\end{equation}
with
\begin{align}
K  &  =\frac{k_{0}\sqrt{s_{0}}E_{d}}{kE_{k/2}W},\label{3.3}\\
k_{0}^{2}  &  =\frac{(s_{0}-m_{K}^{2}-m^{2})^{2}-4m_{K}^{2}m^{2}}{4s_{0}},
\label{3.4}%
\end{align}
where $s_{0}=(W-E_{k/2})^{2}-k^{2}/4$ and the kinematical factor $K$
approaches unity at higher energies. The dotted line is plotted using
Eq.(\ref{3.2}) and is in good agreement with the data. The difference between
this line and the dash-dotted line shows the size of the Fermi motion effect
which comes from the energy dependence of the kaon-nucleon amplitude. This
effect decreases the magnitude of the cross section. Thus the single
scattering impulse approximation underestimates the total cross section at
momenta below $\sim$500 MeV/c. However, as the data of the total cross
sections are rather scattered, all calculations including the full calculation
are consistent with the data at momenta above $\sim$500 MeV/c.%
%TCIMACRO{\FRAME{ftbpFU}{3.0381in}{2.2649in}{0pt}{\Qcb{Total (solid line),
%integrated elastic (dashed line) and integrated inelastic (dash-dotted line)
%$K^{+}d$ cross sections given by our full calculations. The open symbols
%correspond to the Faddeev calculation\cite{ref7,ref8}. The data are from
%Ref.\cite{ref8,ref27}(filled symbols). The circles, triangles and squares
%denote the total, integrated elasic and integrated inelastic cross sections,
%respectively. }}{\Qlb{fig-2}}{fig-2.eps}%
%{\special{ language "Scientific Word";  type "GRAPHIC";
%maintain-aspect-ratio TRUE;  display "USEDEF";  valid_file "F";
%width 3.0381in;  height 2.2649in;  depth 0pt;  original-width 9.3763in;
%original-height 6.9652in;  cropleft "0";  croptop "1";  cropright "1";
%cropbottom "0";  filename 'figure/fig-2.eps';file-properties "XNPEU";}}}%
%BeginExpansion
\begin{figure}
[ptb]
\begin{center}
\includegraphics[
height=2.2649in,
width=3.0381in
]%
{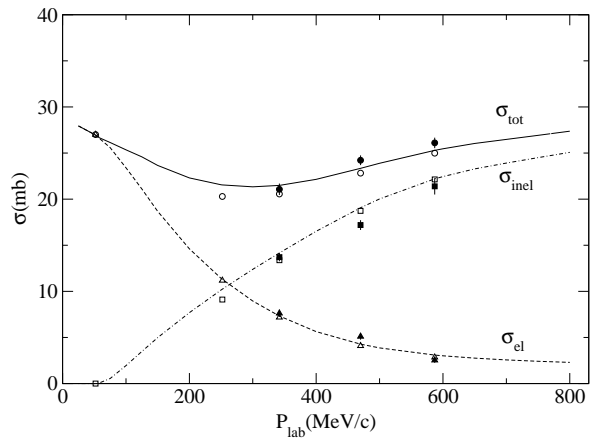}%
\caption{Total (solid line), integrated elastic (dashed line) and integrated
inelastic (dash-dotted line) $K^{+}d$ cross sections given by our full
calculations. The open symbols correspond to the Faddeev
calculation\cite{ref7,ref8}. The data are from Ref.\cite{ref8,ref27}(filled
symbols). The circles, triangles and squares denote the total, integrated
elasic and integrated inelastic cross sections, respectively. }%
\label{fig-2}%
\end{center}
\end{figure}
%EndExpansion

In order to examine the validity of our approach, we compare our full
calculation with the Faddeev calculation by Garcilazo\cite{ref7}. As there are
no data at momenta below 342 MeV/c, we regard the results of the Faddeev
calculation as the data. We use the same kaon-nucleon T matrix used in
Ref.\cite{ref7} but consider only the isospin $I=0$ S-wave nucleon-nucleon
interaction as mentioned above. The results are shown in Fig.2. The solid,
dashed and dash-dotted lines correspond to our full calculations for the
total, integrated elastic and integrated inelastic cross sections,
respectively. The integrated inelastic cross section $\sigma_{inel}$ is given
by $\sigma_{inel}=\sigma_{tot}-\sigma_{el}$. The open circles, open triangles
and open squares correspond to the Faddeev calculations\cite{ref7,ref8} for
the total, integrated elastic and integrated inelastic cross sections,
respectively. The open symbols at the threshold are taken from Fig.1 of
Ref.\cite{ref7}. The filled symbols are the corresponding data measured by
Glasser \textit{et al. }\cite{ref8,ref27}. Our calculations are in good
agreement with the Faddeev calculations as well as the data. The maximum
discrepancy between our calculation and the Faddeev calculation is $6\%$ at
252MeV/c for the total cross section and $8\%$ at 587MeV/c for the elastic
cross section. These results demonstrate that our model, where the optical
potential contains both the first-order and second-order terms, is good enough
to describe these cross sections.%
%TCIMACRO{\FRAME{ftbpFU}{2.9888in}{6.2742in}{0pt}{\Qcb{$K^{+}d$ elastic
%differential cross sections at three incident momenta. The solid line is the
%result of the full calculation. The dashed and dash-dotted line are the
%results of the first order potential and the single scattering impulse
%approximation. The data are from Ref.\cite{ref27}.}}{\Qlb{fig-3}}%
%{m-fig6.eps}{\special{ language "Scientific Word";  type "GRAPHIC";
%maintain-aspect-ratio TRUE;  display "USEDEF";  valid_file "F";
%width 2.9888in;  height 6.2742in;  depth 0pt;  original-width 4.2307in;
%original-height 8.9249in;  cropleft "0";  croptop "1";  cropright "1";
%cropbottom "0";  filename 'figure/m-fig6.eps';file-properties "XNPEU";}}}%
%BeginExpansion
\begin{figure}
[ptb]
\begin{center}
\includegraphics[
height=6.2742in,
width=2.9888in
]%
{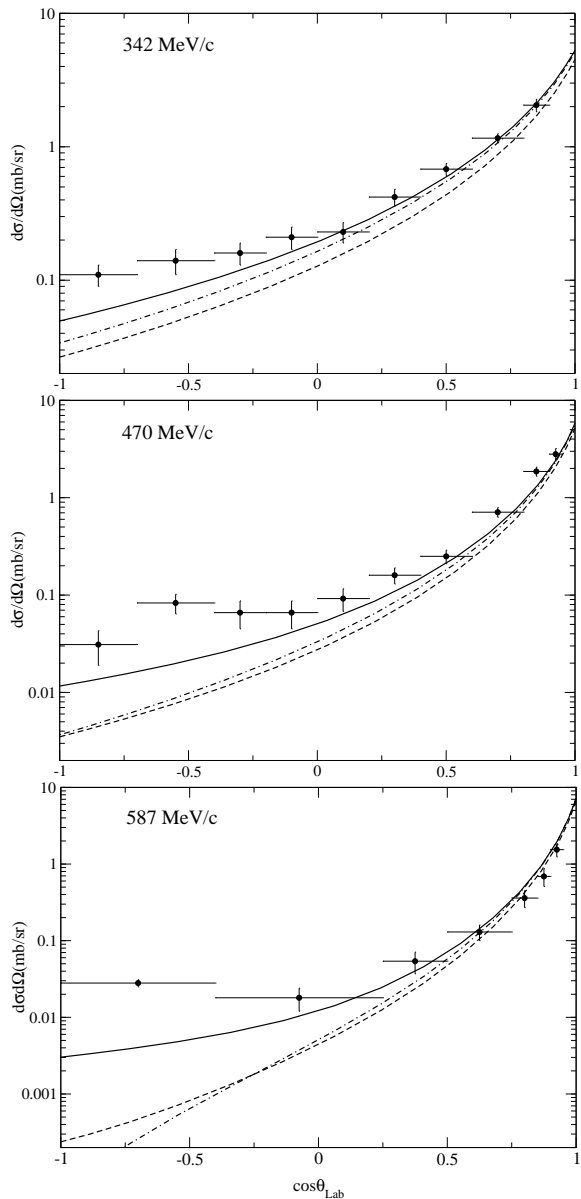}%
\caption{$K^{+}d$ elastic differential cross sections at three incident
momenta. The solid line is the result of the full calculation. The dashed and
dash-dotted line are the results of the first order potential and the single
scattering impulse approximation. The data are from Ref.\cite{ref27}.}%
\label{fig-3}%
\end{center}
\end{figure}
%EndExpansion

To see how our method predicts the elastic differential cross sections, the
calculations at three incident momenta are presented with the data in Fig.3.
The solid, dashed and dash-dotted lines correspond to the full calculation,
the calculation with the first-order optical potential and the single
scattering impulse approximation, respectively. Since the D-wave contribution
in the N-N interaction is disregarded, all calculations at the backward angles
are naturally underestimated. We find that the second-order optical potential
makes the cross section increase and brings it close to the data at backward
angles. For the forward angles, on the other hand, our full calculation is
roughly consistent with the measurement as well as the Faddeev calculation
shown in Figs.2-4 of Ref.\cite{ref7}.
%TCIMACRO{\FRAME{ftbpFU}{3.0346in}{4.305in}{0pt}{\Qcb{Total (top) and
%integrated elastic (bottom) $K^{+}d$ cross sections. The solid and dashed
%lines are the results of the full calculation and the first order optical
%potential, respectively. The dash-dotted line is the result of the single
%scattering impulse approximation. The dash-two-dotted and dotted lines are the
%results of the calculation with the the double scattering effect and the
%modified N-N scattering effect, respectively. See the details in the text. The
%data are the same as Fig.1. }}{\Qlb{fig-4}}{m-fig3.eps}%
%{\special{ language "Scientific Word";  type "GRAPHIC";
%maintain-aspect-ratio TRUE;  display "USEDEF";  valid_file "F";
%width 3.0346in;  height 4.305in;  depth 0pt;  original-width 4.6985in;
%original-height 6.6833in;  cropleft "0";  croptop "1";  cropright "1";
%cropbottom "0";  filename 'figure/m-fig3.eps';file-properties "XNPEU";}}}%
%BeginExpansion
\begin{figure}
[ptb]
\begin{center}
\includegraphics[
height=4.305in,
width=3.0346in
]%
{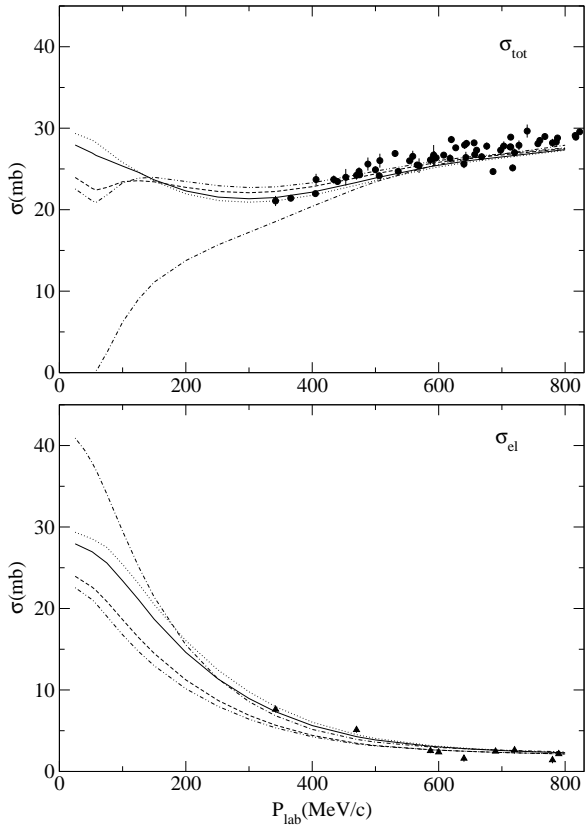}%
\caption{Total (top) and integrated elastic (bottom) $K^{+}d$ cross sections.
The solid and dashed lines are the results of the full calculation and the
first order optical potential, respectively. The dash-dotted line is the
result of the single scattering impulse approximation. The dash-two-dotted and
dotted lines are the results of the calculation with the the double scattering
effect and the modified N-N scattering effect, respectively. See the details
in the text. The data are the same as Fig.1. }%
\label{fig-4}%
\end{center}
\end{figure}
%EndExpansion

In order to see the effects of the multiple scattering, we have calculated the
total and integrated elastic cross sections using several different types of
the optical potential. The results are shown in Fig.4. For reference, the
calculation in the single scattering impulse approximation is plotted as
dash-dotted line. The solid and dashed lines correspond to the results
evaluated with $U^{\left(  1\right)  }+U^{\left(  2\right)  }$ and $U^{\left(
1\right)  }$, respectively. These lines are the same as Fig.1. The
dash-two-dotted and dotted lines are the calculations with $U^{\left(
1\right)  }+U_{d}^{\left(  2\right)  }$ and $U^{\left(  1\right)  }%
+U_{n}^{\left(  2\right)  }-U_{c}^{\left(  2\right)  }$, respectively. The
quantities $U_{d}^{\left(  2\right)  }$ and $U_{n}^{\left(  2\right)  }%
-U_{c}^{\left(  2\right)  }$ represent the effects of the double scattering
and the modified N-N scattering, respectively. From the comparison of the
dashed line with the dash-two-dotted line or the dotted line, we can see the
size of the multiple scattering effects. Although the effects of the double
scattering and the modified N-N scattering are negligible at higher momenta
than 600 MeV/c, they become important with the decreasing of momentum. In the
case of the total cross section shown in the top diagram of Fig.4, the double
scattering term increases the cross section shown by the dashed line and the
modified N-N scattering term decreases it, but at lower momenta than $\sim
$100MeV/c, the effect of the two terms is completely opposite. Near the
threshold, furthermore, all lines except the dash-dotted line agree with the
corresponding lines in the bottom diagram of Fig.4. In the case of the elastic
cross section, on the other hand, the double scattering term decreases the
cross section shown by the dashed line and the modified N-N scattering term
increases it. We also find that the effect of the modified N-N scattering term
is larger than that of the double scattering term.%

%TCIMACRO{\FRAME{ftbpFU}{3.039in}{2.2295in}{0pt}{\Qcb{The same as Fig.1 except
%that the theoretical curves are calculated using the phase shifts of Hyslop
%\QTR{it}{et al.}\cite{ref18}.}}{\Qlb{fig-6}}{fig-5.eps}%
%{\special{ language "Scientific Word";  type "GRAPHIC";
%maintain-aspect-ratio TRUE;  display "USEDEF";  valid_file "F";
%width 3.039in;  height 2.2295in;  depth 0pt;  original-width 9.3763in;
%original-height 6.8554in;  cropleft "0";  croptop "1";  cropright "1";
%cropbottom "0";  filename 'figure/fig-5.eps';file-properties "XNPEU";}}}%
%BeginExpansion
\begin{figure}
[ptb]
\begin{center}
\includegraphics[
height=2.2295in,
width=3.039in
]%
{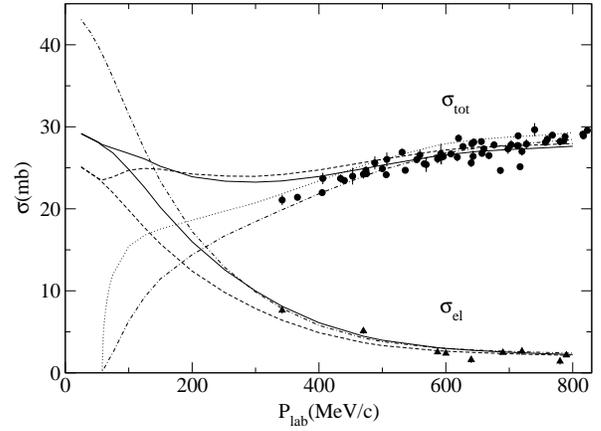}%
\caption{The same as Fig.1 except that the theoretical curves are calculated
using the phase shifts of Hyslop \textit{et al.}\cite{ref18}.}%
\label{fig-6}%
\end{center}
\end{figure}
%EndExpansion

So far we have used the $K^{+}N$ phase shifts of Martin\cite{ref17} in order
to compare our calculation with the Faddeev calculation by
Garcilazo\cite{ref7} and check the validity of our method. Here we examine the
dependence of the one-shell $K^{+}N$ amplitude. In Fig.5, we show the cross
sections calculated by using the phase shifts of Hyslop \textit{et
al}.\cite{ref18}. At $P_{lab}>$ 500 MeV/c, the calculation is in good
agreement with the data. However, at $P_{lab}\approx$ 400 MeV/c, the full
calculation for the total cross section does not agree with the data. Such
discrepancy can be understood from the comparison of the total cross sections
given by Eq.(\ref{3.2}) (see the dotted lines in Fig.1 and Fig.5). The total
cross section represented by the dotted line in Fig.5 is slightly larger than
that in Fig.1, although both of the lines almost agree with the data. Once the
effects of the multiple scattering and the Fermi motion are included in the
calculation, however, this tendency leads to significant difference between
the two calculations of the total cross sections shown in Fig.1 and Fig.5.

\section{\textbf{Conclusions}}

The single scattering impulse approximation is able to describe the $K^{+}d$
data at incident momenta above $\sim$500 MeV/c but explicitly violates the
unitarity below 200MeV/c and furthermore fails to explain the breakup reaction
cross section at the forward kaon scattering angles. Accordingly the single
scattering impulse approximation is not satisfactory for explaining the
$K^{+}d$ scattering. Our purpose was to examine how consistently the optical
potential describes the $K^{+}d$ scattering. This approach does not violate
the unitarity unless an unphysical potential is used. Another advantage of the
optical potential is that the calculation is straightforward compared with the
Faddeev method.

We constructed the optical potential consisting of the first-order and
second-order terms. The second-order optical potential includes the double
scattering term and the modified N-N scattering term. The first-order and
second-order potentials were evaluated without any factorization in the
momentum integration. This potential was used to calculate the total,
integrated elastic and elastic differential cross sections at incident momenta
below 800 MeV/c. We found that our optical potential approach is able to
explain both the Faddeev calculation and the data consistently and especially
the second-order optical potential plays an essential role at low energies.

It was demonstrated in our calculations that the multiple scattering effects
such as the coherent rescattering, the double scattering and the modified N-N
scattering have an important contribution to the cross sections at low
energies in spite of the weak $K^{+}N$ interaction. This importance may be
related to the fact that the wave-length of $K^{+}$ is comparable to the
distance between two nucleons in the deuteron. At low energies, therefore, the
multiple scattering effects should be taken into account when one extracts the
on-shell $K^{+}N$ amplitudes from the $K^{+}d$ data. This was confirmed
through the comparison of the calculations obtained using two kinds of
$K^{+}N$ phase shifts.

It is interesting to examine whether our approach is applicable for other
reactions arising from the strong elementary interaction such as a $\pi d$
scattering and a $K^{-}d$ scattering. Furthermore our optical potential
approach could be used to study the effect of the pentaquark resonance
$\Theta(1540)$ in the $K^{+}d$ scattering and especially learn how the effects
of the multiple scattering affect the suppression of the resonance peak in the
$K^{+}d$ cross sections.

Our optical potential has been derived from the Watson multiple scattering
theory. Similarly, the optical potential up to second order can be formulated
based on the KMT multiple scattering theory as shown in Appendix. We have
numerically tested the difference between two approaches. Within the
first-order optical potential approach, the result by the Watson formulation
does not agree with that by the KMT formulation. In the latter calculation,
the integrated elastic cross section becomes larger than the total cross
section at low momenta. This does not mean that the imaginary part of the KMT
first-order optical potential is positive at low energies, but this is due to
the factor $\frac{A-1}{A}$ appeared in the KMT formulation. When the
second-order optical potential is taken into account, however, the two
approaches give almost the identical cross sections. It was found from our
numerical estimate that the difference was a few percent or less except near
the threshold. Therefore our conclusions in this paper are not changed by
which formulation is chosen.

\begin{acknowledgments}
The author would like to thank M.Hirata for useful discussions.
\end{acknowledgments}

\appendix{}

\section*{Appendix}

We will derive the optical potential from the KMT multiple scattering theory.
The transition amplitude $T$ of Eq.(\ref{2.1}) is rewritten as%
\begin{equation}
T=\sum_{i=1}^{A}T_{i}, \tag{A1}\label{a1}%
\end{equation}
where%
\begin{align}
T_{i}  &  =\tau_{i}+\tau_{i}\frac{\mathcal{A}}{e}\sum_{j\neq i}T_{j}%
,\tag{A2}\label{a2}\\
\tau_{i}  &  =v_{i}+v_{i}\frac{\mathcal{A}}{e}\tau_{i}. \tag{A3}\label{a3}%
\end{align}
In order to get the KMT optical potential $U_{KMT}$, we introduce the operator
$\tilde{U}_{i}$ defined by
\begin{equation}
T_{i}=\tilde{U}_{i}+\tilde{U}_{i}\frac{P}{e}\sum_{j\neq i}T_{j}.
\tag{A4}\label{a4}%
\end{equation}
In the KMT formulation, all equations are derived in the antisymmetric
subspace of the Hilbert space. Consequently, the operators of $\tau_{i}$,
$\tilde{U}_{i}$ and $T_{i}$ are independent of $i$. With the help of the
relation $\sum_{j\neq i}T_{j}=\frac{A-1}{A}T$, Eq.(\ref{a4}) can be expressed
as%
\begin{equation}
T^{^{\prime}}=U_{KMT}+U_{KMT}\frac{P}{e}T^{^{\prime}}, \tag{A5}\label{a6}%
\end{equation}
where%
\begin{align}
T^{^{\prime}}  &  =\frac{A-1}{A}T,\tag{A6}\label{a7}\\
U_{KMT}  &  =\frac{A-1}{A}\sum_{i=1}^{A}\tilde{U}_{i}. \tag{A7}\label{a9}%
\end{align}
Similarly, Eq.(\ref{a2}) is written as%
\begin{equation}
T^{^{\prime}}=\tau_{KMT}+\tau_{KMT}\frac{\mathcal{A}}{e}T^{^{\prime}},
\tag{A8}\label{a9-1}%
\end{equation}
where%
\begin{equation}
\tau_{KMT}=\frac{A-1}{A}\sum_{i=1}^{A}\tau_{i}. \tag{A9}\label{a9-2}%
\end{equation}
The scattering operator $T$ is obtained by solving Eq.(\ref{a6}), if the
optical potential $U_{KMT}$ is given. With the help of Eqs.(\ref{a6}) and
(\ref{a9-1}), \ the operator $U_{KMT}$ can be written in terms of $\tau_{i}$
as%
\begin{align}
U_{KMT}  &  =\tau_{KMT}+\tau_{KMT}\frac{Q}{e}U_{KMT}\tag{A10}\label{a11}\\
&  =\frac{A-1}{A}\sum_{i=1}^{A}\tau_{i}+\frac{A-1}{A}\sum_{i=1}^{A}\tau
_{i}\frac{Q}{e}U_{KMT}. \tag{A11}\label{a12}%
\end{align}

Now we build the first-order and second-order optical potentials using
Eq.(\ref{a12}). The operator $\tau_{i}$ can be written in terms of $t_{i}$ as
Eq.(\ref{2.9-1}) in Sec.2. Here the operator $t_{i}$ is defined by
Eq.(\ref{2.8}). By substituting Eq.(\ref{2.9-1}) into Eq.(\ref{a12}), the
operator $U_{KMT}$ can be expressed in terms of $t_{i}$ as%
\begin{equation}
U_{KMT}=\frac{A-1}{A}(\sum_{i}t_{i}+\sum_{i\neq j}t_{i}\frac{1}{e}t_{j}%
-\sum_{i\neq j}t_{i}\frac{P}{e}t_{j}+\cdots).\tag{A12}\label{a13}%
\end{equation}
We note that an intermediate state in the operators $U_{KMT}$ and $t_{i}$ is
either an antisymmetric state or a non-antisymmetric state. Eq.(\ref{a13})
displays the KMT optical potential to second order in $t_{i}$, which
corresponds to the Watson optical potential of Eq.(\ref{2.11}). Now we
consider the optical potential $U_{KMT}$ used in the calculation of the
$K^{+}d$ scattering. Using the same procedure mentioned in Sec.2, we obtain
the final expression%
\begin{equation}
U_{KMT}=U_{KMT}^{(1)}+U_{KMT}^{(2)}+\cdots,\tag{A13}\label{a14-1}%
\end{equation}
with
\begin{align}
U_{KMT}^{(1)} &  =\frac{1}{2}U^{(1)},\tag{A14}\label{a15}\\
U_{KMT}^{(2)} &  =\frac{1}{2}(U^{(2)}+\sum_{i}t_{i}^{free}\frac{P}{e}%
t_{i}^{free}),\tag{A15}\label{a16}%
\end{align}
where the operators $U^{(1)}$, $U^{(2)}$ and $t_{i}^{free}$ are defined in Sec.2.

\end{document}